\documentclass[11pt,a4paper,twocolumn]{article}

\usepackage[latin1]{inputenc}
\usepackage{amsmath}
\usepackage{amsfonts}
\usepackage{amssymb}
\usepackage[left=1.5cm,right=1.5cm,top=1.5cm,bottom=1.5cm]{geometry}
\usepackage{graphicx}
\usepackage{bm}
\usepackage{xcolor}
\usepackage{authblk}

\newif\ifcolourchanges

\newcommand{\kT}{k_{\mathrm{B}}T}

\newcommand{\vecr}{\textbf{r}}

\newcommand{\Ncol}{N_\mathrm{col}}
\newcommand{\Npoly}{N_\mathrm{poly}}
\newcommand{\Nmono}{N_\mathrm{mono}}
\newcommand{\Ntad}{N_\mathrm{tad}}
\newcommand{\Ntail}{N_\mathrm{tail}}

\newcommand{\Rcol}{R_\mathrm{col}}
\newcommand{\Rmono}{R_\mathrm{mono}}
\newcommand{\Rhead}{R_\mathrm{head}}
\newcommand{\Rtail}{R_\mathrm{tail}}

\newcommand{\phic}{\phi_\mathrm{col}}
\newcommand{\phip}{\phi_\mathrm{poly}}
\newcommand{\phit}{\phi_\mathrm{tad}}
\newcommand{\phis}{\phi_\mathrm{s}}

\newcommand{\Sconfig}{S_m}
\newcommand{\Sconform}{S_c}

\begin{document}

\title{Surfactancy in a tadpole model of proteins}

\author[1]{O. T. Dyer}
\author[1]{R. C. Ball}
\affil[1]{%
 Department of Physics, University of Warwick, Coventry CV4 7AL, United Kingdom
}%
\affil[]{oliver.dyer@warwick.ac.uk, R.C.Ball@warwick.ac.uk}


\maketitle

\begin{abstract}
We model the environment of eukaryotic nuclei by representing macromolecules by only their entropic properties, with globular molecules represented by spherical colloids and flexible molecules by polymers.
We put particular focus on proteins with both globular and intrinsically disordered regions, which we represent with `tadpole' constructed by grafting single polymers and colloids together.
In Monte Carlo simulations we find these tadpoles support phase separation via depletion flocculation, and demonstrate several surfactant behaviours, including being found preferentially at interfaces and forming micelles in single phase solution.
Furthermore, the model parameters can be tuned to give a tadpole a preference for either bulk phase.
However, we find entropy too weak to drive these behaviours by itself at likely biological concentrations.
\end{abstract}


\section{Introduction}

Liquid-liquid phase separation (LLPS) has been identified as a mechanism for eukaryotic cells to organise their constituent biomolecules without needing the lipid bilayer membranes used by conventional organelles \cite{Hyman2014,Alberti2017,Shin2017}.
The resulting droplets are sometimes called `membraneless organelles' \cite{Sawyer2019}. 
The best and longest known example of these is the nucleolus, which is involved in the production of ribosomes \cite{Olson2015}, but more recent work has identified a variety of similar - albeit smaller - liquid aggregates inside both the nucleus and the cell cytoplasm \cite{Sawyer2019}.
Their locations, functions and compositions vary, but they frequently feature proteins with long sections of polypeptide chain that remain flexible instead of folding into rigid secondary and tertiary structures \cite{Shin2017,Galganski2017,Boehning2018,Boija2018}.
These sections are called intrinsically disordered regions (IDRs) and can be found anywhere along the polypeptide chain.
In this article we will model IDRs at either the carboxy terminal domain (CTD) or amino terminal domain, leading to proteins with a flexible chain attached to a single globular region.

The role of these IDRs in LLPS remains uncertain, with most work to date focusing on weak attractive interactions between IDRs, especially those between charge dipoles, aromatic groups and net charges in amino acid residues \cite{Brangwynne2015}.
The idea being that having longer IDRs provides more weak interactions with which to hold the droplet together.

These energetic considerations are not alone in driving phase behaviour, however, as entropy also plays a role.
This has several contributions: entropy of mixing drives systems to mix, and thus hinders LLPS; conformational entropy of IDRs multiplies the number of microstates utilising the chemistry above \cite{Fuxreiter2018}; and even without chemistry the conformational entropy introduces crowding effects that give IDRs an effective repulsion from globular molecules.
Crowding effects, which are the focus of this work, are not new to cell biology.
Studies have included the organisation of DNA at different length scales \cite{Marenduzzo2006,Brackley2013,CanalsHamann2013,Cook2018,Oh2018}, the influence of crowding on transcription rates \cite{Matsuda2014} and on LLPS \cite{Marenduzzo2006,Cho2012,Bakshi2015,Kaur2019,Andre2020}.
However, to our knowledge purely replusive models of proteins with both IDRs and globular regions have not yet been studied.

Physics can provide insights into such systems by drawing analogy to polymer-colloid systems, which are characterised by the ratio of polymer radius of gyration and colloid radius $q=R_g / \Rcol$.
Behaviour in the `colloid limit', $q\ll 1$, is very well established, with mixtures of non-adsorbing polymers and colloids separating via depletion flocculation \cite{Asakura1954,Lekkerkerker1992,Ilett1995}.
The story in the `protein limit', $q > 1$, is similar, but with the key difference that phase separation is less sensitive to $R_g$ than the correlation length of the polymer mesh \cite{Bolhuis2003}.
Recent work has argued similar mechanisms can play a role in biology where it can separate nuclear transport receptors in nuclear pores \cite{Davis2022}.

Interesting behaviour also arises when the polymers are grafted to proteins.
In the colloid limit, it is common practice to coat the colloids with polymers, stabilising them against the van der Waals forces driving flocculation by exploiting the entropic repulsion of the polymers \cite{Tadros2013}.
In the protein limit, polymer-colloid grafts are often called grafted nanoparticles (GNPs), and have been studied extensively in the last two decades \cite{Zhang2003,Iacovella2005,Iacovella2008,Iacovella2011,Meli2009,Kumar2013,Marson2015,Kumar2017}.
This work has focused largely on materials science, usually aiming to improve the properties of nanocomposites \cite{Kumar2013} or using specific interactions of the GNPs in the design of materials \cite{Marson2015}.
Research into GNPs with only one grafted polymer - an appropriate analogue to proteins with a single IDR - has been comparatively sparse \cite{Zhang2003,Iacovella2005,Iacovella2008,Marson2015}.

Similarly to the biological LLPS literature, we are unaware of any work that has sought to quantify the importance of entropy in GNP behaviour.
Hence, by modeling proteins with IDRs as single-chain GNPs with no attractive interactions, this article simultaneously fills gaps in both fields.

Finally, we recognise solvent entropy can also play a role via hydration effects and has been shown to be important in protein aggregation, even for proteins without IDRs \cite{Matsarskaia2016,Sahoo2022,Bianco2019,Bianco2020,Park2020}. 
In this work we only consider explicitly the entropic properties of the macromolecules, 
with solvent entropy folded into net local solute interactions which we assume to be fully repulsive.

\begin{figure}
    \centering
    \includegraphics[width=0.9\linewidth]{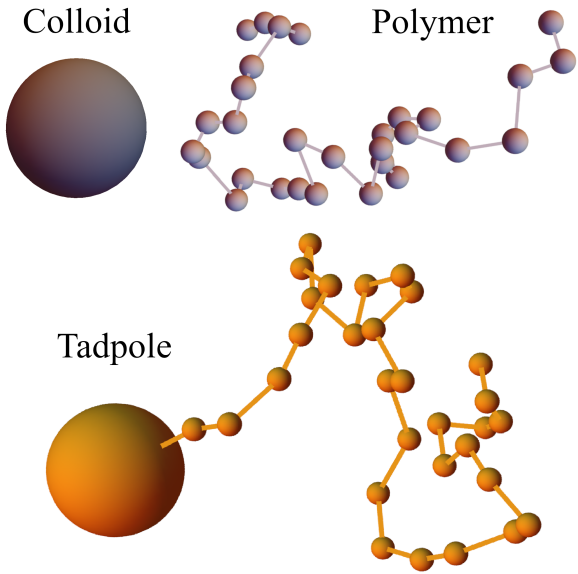}
    \caption{Examples of our model molecules. 
        Note they are 3 dimensional, so apparent overlaps are not real.
        The tadpole's `head' (large bead) is equivalent to a colloid, while the `tail' is equivalent to a polymer. 
        All beads are hard spheres and interact with potentials in Eqs.~\eqref{Eq: hard sphere potential} and \eqref{Eq: bond potential}.}
    \label{Fig: tadpole model}
\end{figure}

\section{Methods}
\label{Sec: Methods}

To study the role of IDRs' entropy in LLPS, we run dynamic Monte Carlo simulations in continuous, 3-dimensional, periodic boxes.
The box dimensions are $L_x \times L_y \times L_z$, where we always set $L_x = L_y$ and take $L_z$ as the box's longest dimension.
These dimensions will vary between sections, where they are stated.

We coarse grain macromolecules into collections of spheres where the $i$th particle has hard core radius $R_i$.
Since we focus solely on their entropic properties, these particles experience no attractions and repel each other with the hard sphere potential 
\begin{equation}
    U_{\mathrm{HS}}(r_{ij}) = \left\lbrace 
        \begin{array}{ll}
            \infty & r_{ij} < R_{i} + R_{j} \\
            0 & \mathrm{otherwise}
        \end{array}
    \right.
\label{Eq: hard sphere potential}
\end{equation}
when particles $i$ and $j$ are distance $r_{ij}$ apart.

Spherical particles are joined together into the 3 molecule species depicted in Fig.~\ref{Fig: tadpole model}.
Colloids are single large beads with radius $\Rcol$ representing low-entropy (globular) macromolecules.
Polymers are flexible, linear chains of $\Nmono$ small beads, each with radius $\Rmono$, that represent molecules with high conformational entropy.
Finally, `tadpoles' represent IDR-containing proteins with a globular region attached to a single IDR.
These two regions are represented by a large `head' bead with radius $\Rhead$, and a `tail' chain of $\Ntail$ small beads with radii $\Rtail$.
To reduce the number of parameters in this work, we set $\Rhead = \Rcol$ and $\Rtail = \Rmono$, so a tadpole can be viewed as a colloid grafted to a polymer when $\Ntail = \Nmono$.
This helps minimise the number of parameters used in this work, but limits our ability to make quantitative comparisons to experiments.

All intra-molecular bonds are enforced with the hard square well potential
\begin{equation}
    U_{\mathrm{bond}}(r_{i,i+1}) = \left\lbrace 
        \begin{array}{ll}
            0 & r_{i,i+1} < R_{i} + R_{i+1} + 3\Rtail \\
            \infty & \mathrm{otherwise}
        \end{array}
    \right. ,
\label{Eq: bond potential}
\end{equation}
which acts alongside $U_{\mathrm{HS}}$.

These molecules are placed in an implicit solvent and do not experience any hydrodynamic effects beyond their diffusivity being set by the usual Stokes-Einstein expression $D_i = \kT /(6\pi \eta R_{i})$.
Rather than specify the viscosity $\eta$ explicitly, we define the time unit by the time taken for small particles to diffuse their own size, as listed in Table \ref{Table: parameters}.
Similarly, we use the thermal energy $\kT$ as our energy unit.

This diffusivity is achieved by displacing particles with the noise term used in the Brownian dynamics algorithm \cite{Allen2017},
\begin{equation}
    \vecr_i (t+\delta t) = \vecr_i (t) +
        \sqrt{2 D_i\, \delta t} \textbf{W},
\end{equation}
where $\textbf{W}$ is a Gaussian-distributed random vector with zero mean and unit variance in all 3 directions.  
The conservative interactions are implemented with a Metropolis acceptance test, passing with probability
\begin{equation}
    P_\mathrm{accept} = \mathrm{min}\left[1,\exp(-\Delta U / \kT)\right].
\end{equation}
With our hard systems, this means only accepting displacements if the particle's final energy is zero.
To maintain a high acceptance rate in this hard environment we set the time increment $\delta t$ such that $\sqrt{2 D_i\, \delta t} < \Rtail$.

The values of parameters used in this work are listed in Table \ref{Table: parameters}.
Our choices for particle sizes lead to $q\approx 2.3$ when $\Ntail=30$, which approximates the ratio for human RNA polymerase II's globular region and the flexible CTD on its largest subunit \cite{Davis2002,Boehning2018}.

Our criterion for equilibration is described in Appendix \ref{App: verifying equilibration}.
Except where otherwise stated, our quantitative results average over the last 10 system configurations in each simulation, each separated in time by at least $3\times10^{4}\tau$ to allow for local re-equilibration.  
We then further averaged over at least 7 independent simulation runs for each system, giving us a minimum of 70 configurations contributing to each data point.

\begin{table}
\caption{\label{Table: parameters} 
Parameter values in this work, with code units specified in the top rows to help ground our results in their biological context. 
Where multiple values are used, their ranges are specified. }
\begin{tabular}{cc}
Parameter & Value in simulation \\
\hline
Length unit, $\ell$ & 12.5nm \\
Time unit, $\tau$ & $\pi \eta \Rtail^3 /\kT$\\
Energy unit, $\varepsilon$ & $\kT$ \\ 
\hline
$\Rcol/\ell$, $\Rhead/\ell$ & 0.44 \\ 
$\Rmono/\ell$, $\Rtail/\ell$ & 0.0908\\
$\Ncol$ & 0-376\\
$\Npoly$ & 0-188 \\
$\Nmono$ & 30 \\
$\Ntad$ & 0-188 \\
$\Ntail$ & 15-30 \\
$\delta t / \tau$ & 0.4
\end{tabular}
\end{table}

\section{Entropic phase separation}

Before showing the results of these simulations, it is helpful to explain the physics driving phase behaviour in our minimal model.
By only considering hard interactions, all allowed states have the same energy and the temperature does not affect equilibrium states (although it does still play a role in dynamics).
Free energy differences can therefore be written only in terms of the entropy:
\begin{equation}
    dF = -TdS = -T d\Sconfig - T d\Sconform.
\end{equation}
Here we have split the entropy of mixing ($\Sconfig$) and conformational entropy ($\Sconform$), characterising arrangements of molecule centres of mass and intra-molecule arrangements respectively.

This split is relevant for crowding effects where flexible chains are obstructed by nearby rigid bodies, reducing the number of available conformations of the chain, and in turn leading to a significant reduction in $\Sconform$.
The archetypal system for this is a mixture of colloids and polymers, in which the two species can phase separate via depletion flocculation, where the dominance of $\Sconform$ over $\Sconfig$ drives the colloids to cluster to minimise the contact of the two species, thus allowing the polymers to adopt the maximum number of conformations \cite{Asakura1954}.

The expectation is that this remains significant in our systems with tadpoles present, and hence that proteins may have evolved IDRs to exploit it.

\section{Results}

\subsection{Qualitative phase behaviour}
\label{Sec: qualitative behaviour}

\begin{figure*}
    \centering
    \includegraphics[width=\linewidth]{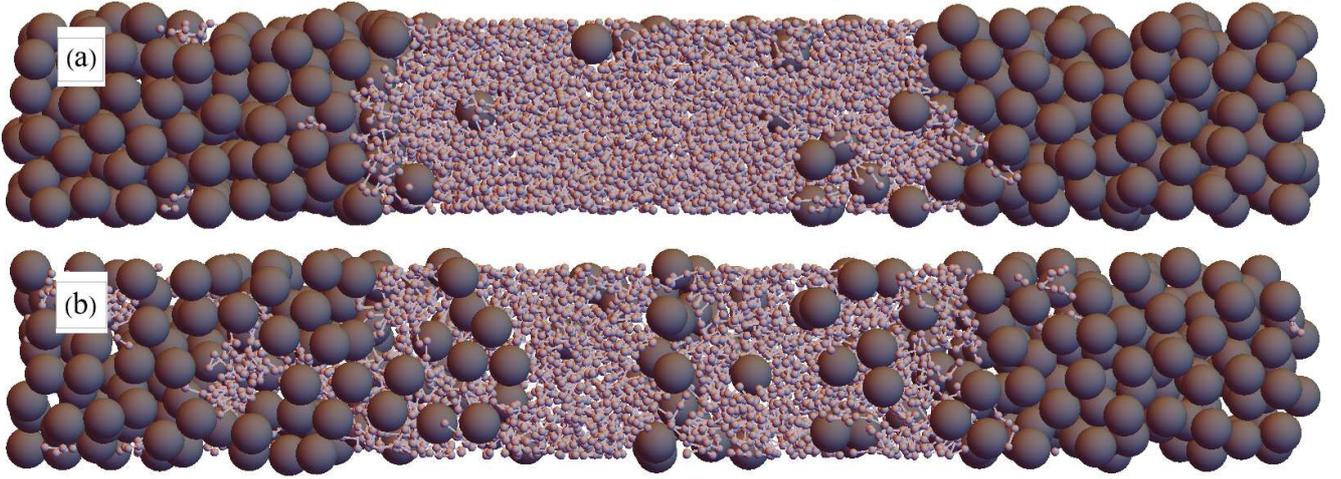}
    \caption{Snapshots of equilibrated polymer + colloid systems.
    Both systems have box dimensions $4\times4\times27$.
    (a): $\Ncol=374$, $\Npoly=188$.
    (b): $\Ncol=328$, $\Npoly=126$.}
    \label{Fig: CP snapshots}
\end{figure*}

We first confirm depletion flocculation occurs in our simulations by simulating systems with only polymers and colloids present.
For this we used highly stretched periodic boxes, with dimensions $4\times 4\times 27$, that produce well-defined planar interfaces perpendicular to the $z$-axis, and whose widths are much smaller than the bulk phases.
The transverse dimensions of the box are large enough to prevent molecules interacting with their own periodic images, but are quite small to keep the computational cost of simulations manageable.
Such systems are shown in Fig.~\ref{Fig: CP snapshots}, in which the denser system in part (a) shows a clear separation into polymer- and colloid-rich phases, as expected.
The interfaces between these phases are also predictably close to planar and perpendicular to the box's long axis.

Reducing the system density allows for a greater concentration of the minority component in each phase, which is particularly noticeable by the number of colloids in the polymer-rich phase in Fig.~\ref{Fig: CP snapshots}(b). 
Reducing the densities further ultimately reaches fully mixed states, and this will be analysed more quantitatively in Sec.~\ref{Sec: Phase diagrams}.

Next, we add tadpoles to our systems, utilising the properties of our model molecules to make the replacement polymer + colloid $\rightarrow$ tadpole.
In this way the total number of large particles (colloids and tadpole heads) is kept constant, as is the number of small particles (in polymers and tadpole tails).
This process can be viewed as the addition of $\Ntad$ bonds fusing polymers to colloids.
Starting from the systems in Fig.~\ref{Fig: CP snapshots}, the concentrations are high enough that nonlinear terms in the osmotic pressure dominate, so the osmotic pressure is close to constant in each series of systems thus created, despite the number of molecules being reduced by $\Ntad$ from the initial polymer + colloid system.

\begin{figure*}
    \centering
    \includegraphics[width=\linewidth]{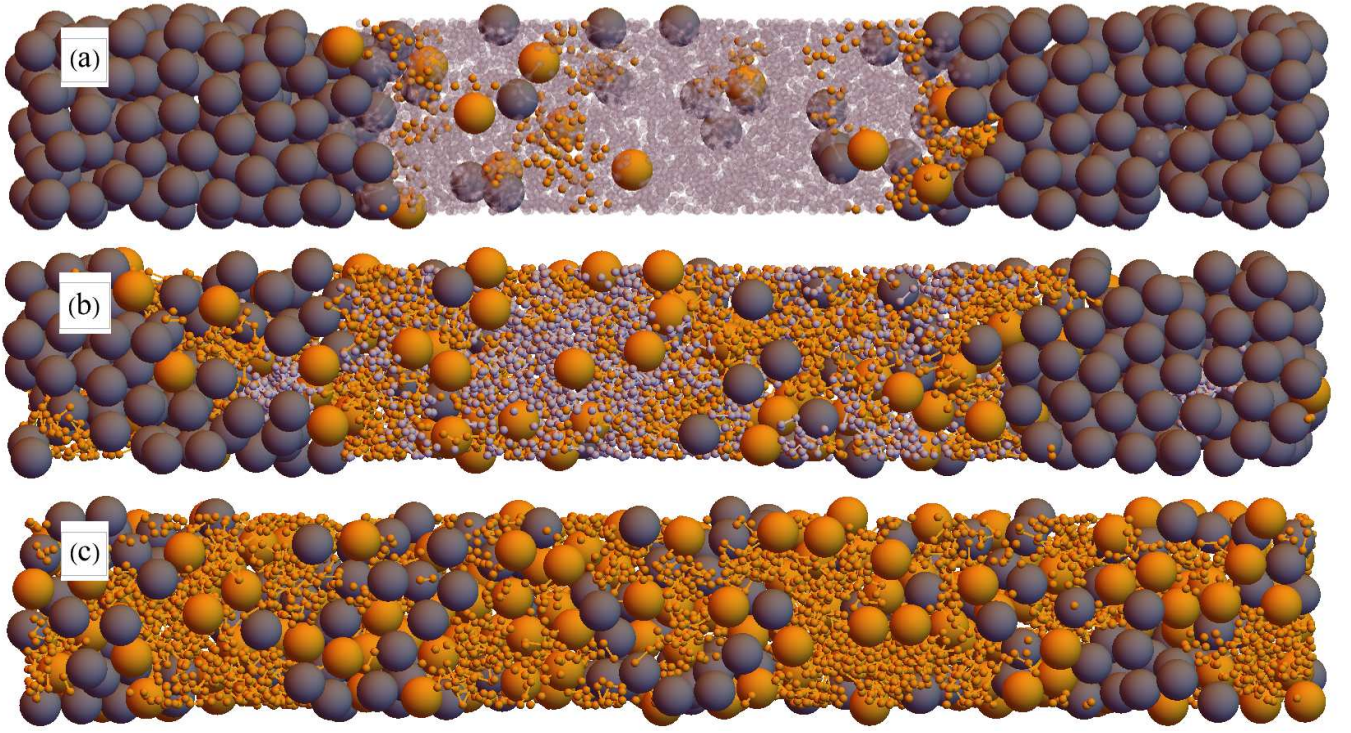}
    \caption{Snapshots of systems in the series replacing colloids and polymers with tadpoles (orange), starting from the system shown in Fig.~\ref{Fig: CP snapshots}(a). 
    The number of tadpoles added is (a) 14, (b) 94, (c) 188 (all polymers replaced).
    The opacity of polymers in part (a) is reduced to improve visibility of tadpoles in the polymer phase and at the interfaces.
    As in Fig.~\ref{Fig: CP snapshots}, the box dimensions are $4\times4\times27$.}
    \label{Fig: CTP snapshots}
\end{figure*}

\begin{figure}
\includegraphics[width=\linewidth]{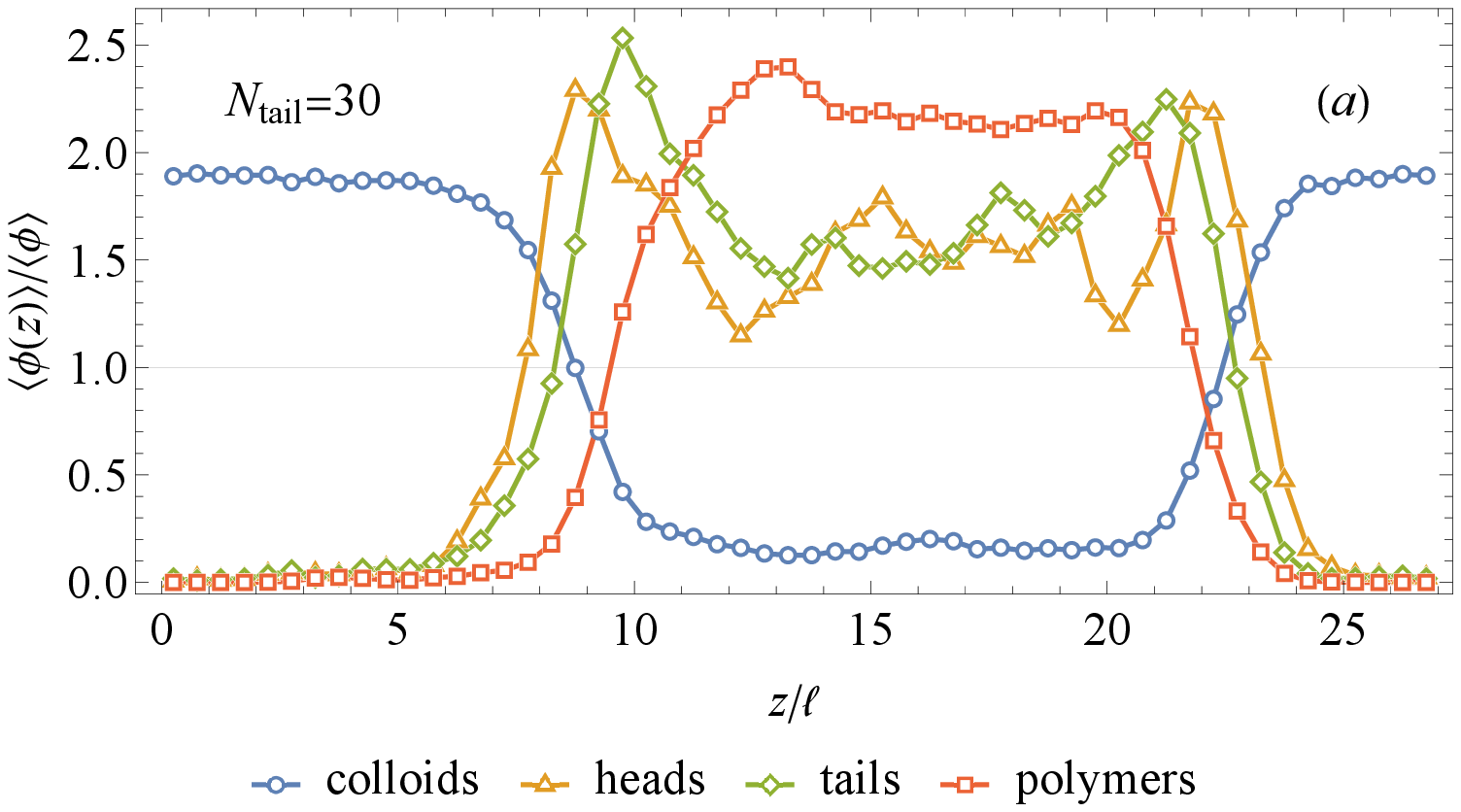}
\includegraphics[width=\linewidth]{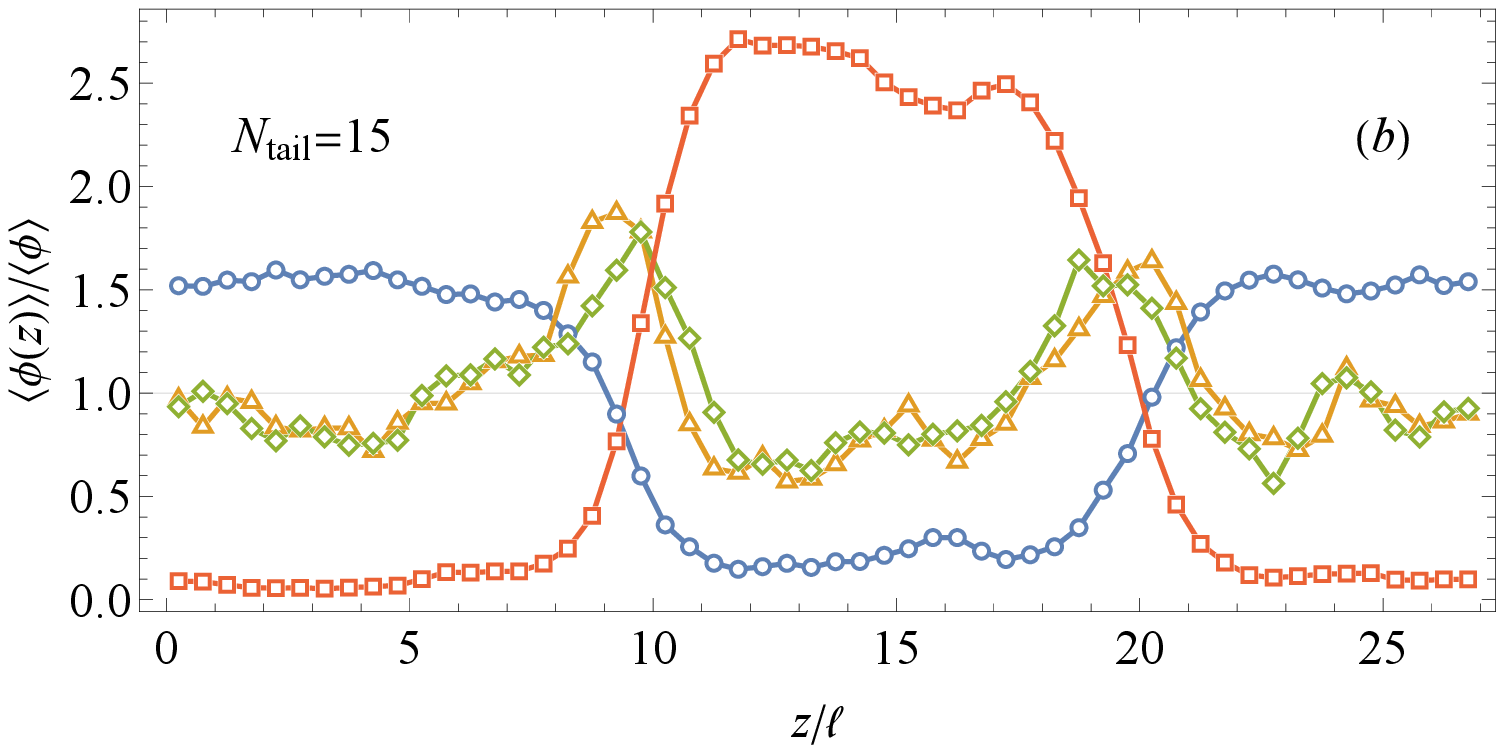}
\caption{Plots of species volume fractions as a function of position along the simulation box, scaled by their global values.
Both plots are taken from systems with $\Ncol=329$, $\Npoly=141$ and $\Ntad=47$, with $\Ntail$ as indicated, and are averages over 80 simulation snapshots, each translated to align their interfaces.}
\label{Fig: phi vs x}
\end{figure}

Snapshots from such a series of simulations starting from Fig.~\ref{Fig: CP snapshots}(a) are shown in Fig.~\ref{Fig: CTP snapshots}.
Part (a) demonstrates that adding a small number of tadpoles this way has minimal impact on the volumes of the two bulk phases, and that of the two they prefer the polymer-rich phase.
Fig.~\ref{Fig: phi vs x}(a), which averages volume fractions in systems that have fused together 47 tadpoles, quantitatively shows the tadpole density drops to almost zero in the colloid-rich phase.

Furthermore, in Fig.~\ref{Fig: CTP snapshots}(a) there appears to be a higher density of tadpoles at the interfaces than in the bulk.
This would make intuitive sense as their heads and tails each prefer different phases when not bonded together, and hence one might expect tadpoles to want to straddle the interface to satisfy the preference of both their head and tail.
This surfactant-like behaviour is confirmed quantitatively in Fig.~\ref{Fig: phi vs x}(a).
Here, the tadpole density peaks at the interfaces, and close inspection finds the tadpoles tend to orient themselves so their heads and tails reside in the colloid and polymer phases respectively, as seen by the peaks of the head density being closer to the colloid phase than the corresponding peaks in tail density.

When half of the polymers have been fused into tadpoles, shown in Fig.~\ref{Fig: CTP snapshots}(b), the two phases remain separated but the polymer-rich phase expands to accommodate the additional volume of the tadpole heads.
The interfaces have become more jagged, suggesting a reduction in interface tension (a point we will return to in Sec.~\ref{Sec: Interface tension}), which is expected if the tadpoles are acting as surfactants.
The tadpoles do not perfectly line the interfaces, however, which is indicative of the limited strength of the entropic `forces' holding them there.

\begin{figure}
    \centering
    \includegraphics[width=0.9\linewidth]{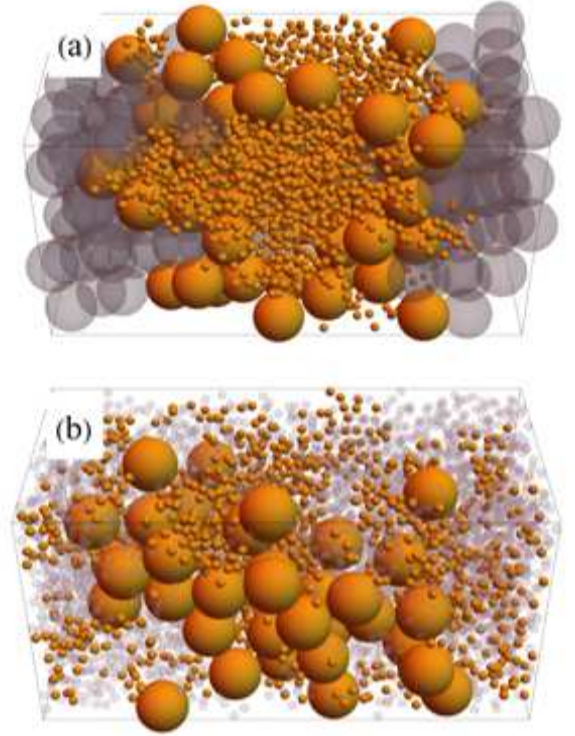}
    \caption{Example systems in which tadpoles form micelles, (a), and inverted micelles, (b).
    (a): tadpole + colloid system with $\Ntad = 56, \Ncol = 112$.
    (b): tadpole + polymer system with $\Ntad = 42, \Npoly = 126$. 
    The opacity of colloids and polymers has been reduced to help see the tadpoles.
    Both systems have box dimensions $4\times4\times8$.}
    \label{Fig: micelles}
\end{figure}

Adding yet more tadpoles, ultimately to the complete removal of polymers, yields the microphase separation seen Fig.~\ref{Fig: CTP snapshots}(c).
In this case we find bicontinuous phases of tadpole tails and tadpole heads + colloids.
By changing species concentrations, we can coax the tadpoles into forming micelles. 
Snapshots of these in shorter simulation boxes, $4\times 4\times 8$, are shown in Fig.~\ref{Fig: micelles}.
In part (a), colloid density has been increased from the bicontinous phase, producing a cylindrical micelle with tadpole tails inwards, and whose axis is along the $x$-axis (front-to-back).
For part (b), replacing colloids with polymers inverts the micelle so that the tadpole tails now point outwards, leaving a rod-like cluster of heads at its centre.
These microphases are commonly produced by surfactant molecules \cite{Marques2013}, adding further evidence that tadpoles act as entropic surfactants.

\subsection{Phase diagrams}
\label{Sec: Phase diagrams}

\begin{figure*}
    \centering
    \includegraphics[width=0.47\linewidth]{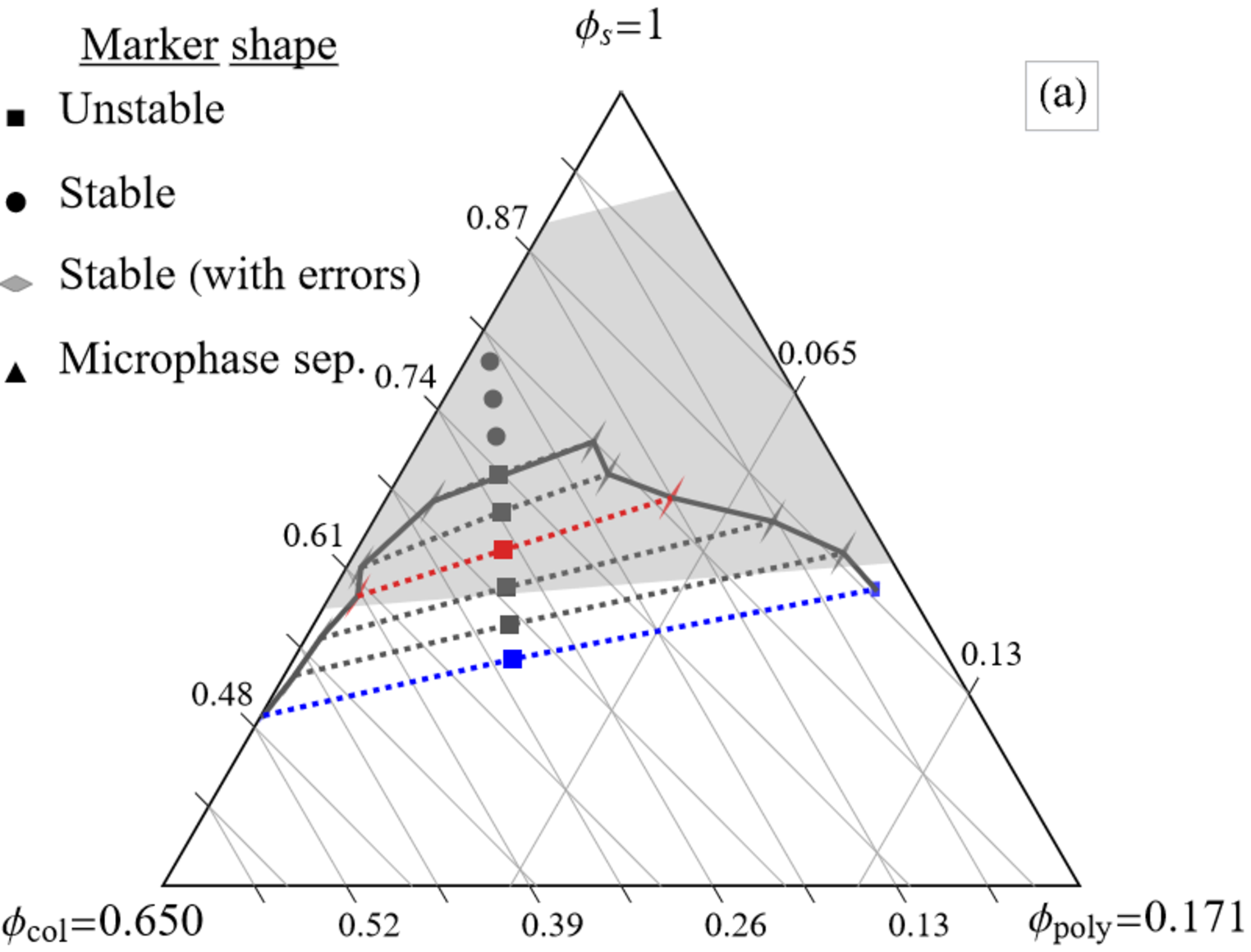}
    \includegraphics[width=0.51\linewidth]{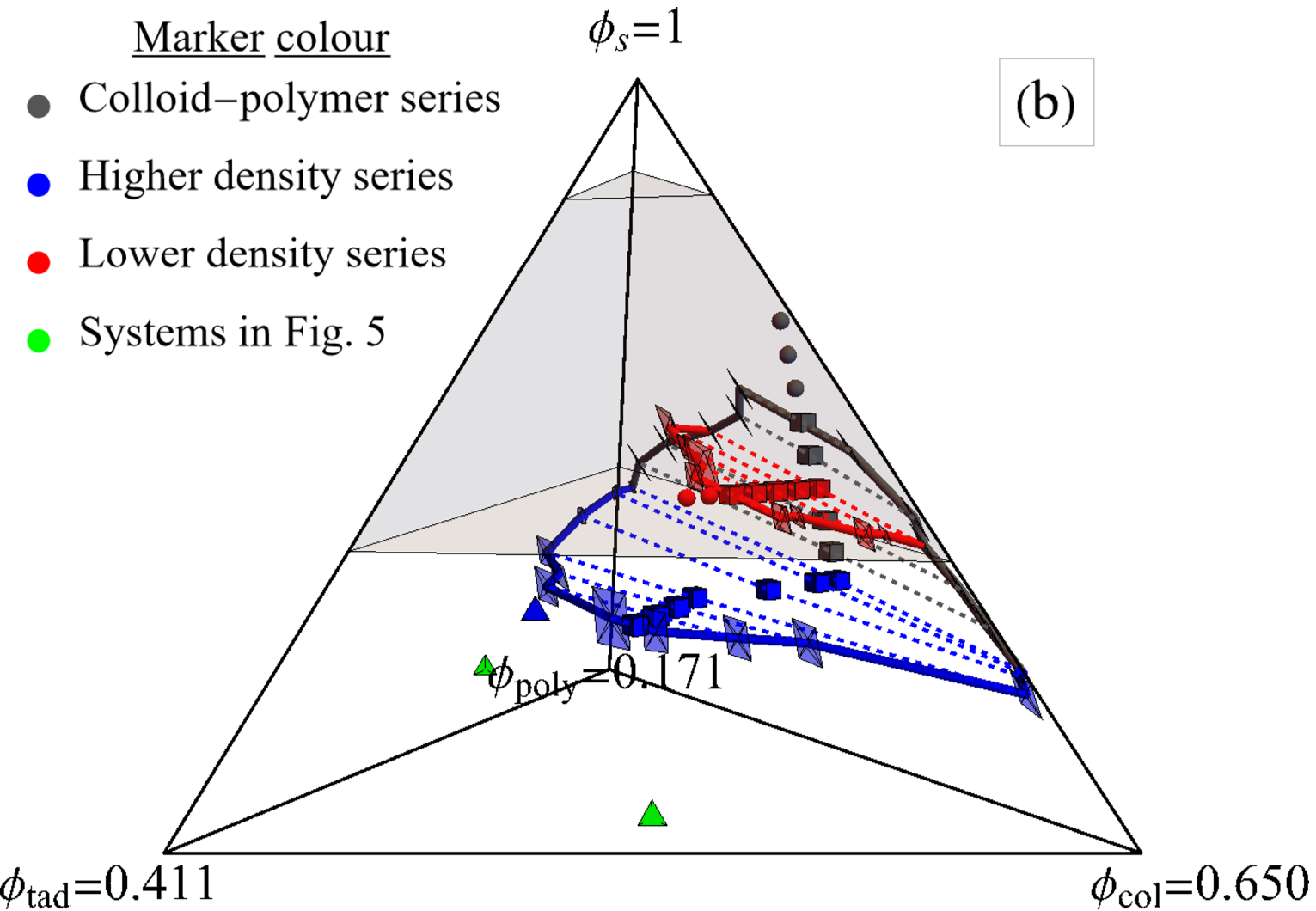}
    \includegraphics[width=0.49\linewidth]{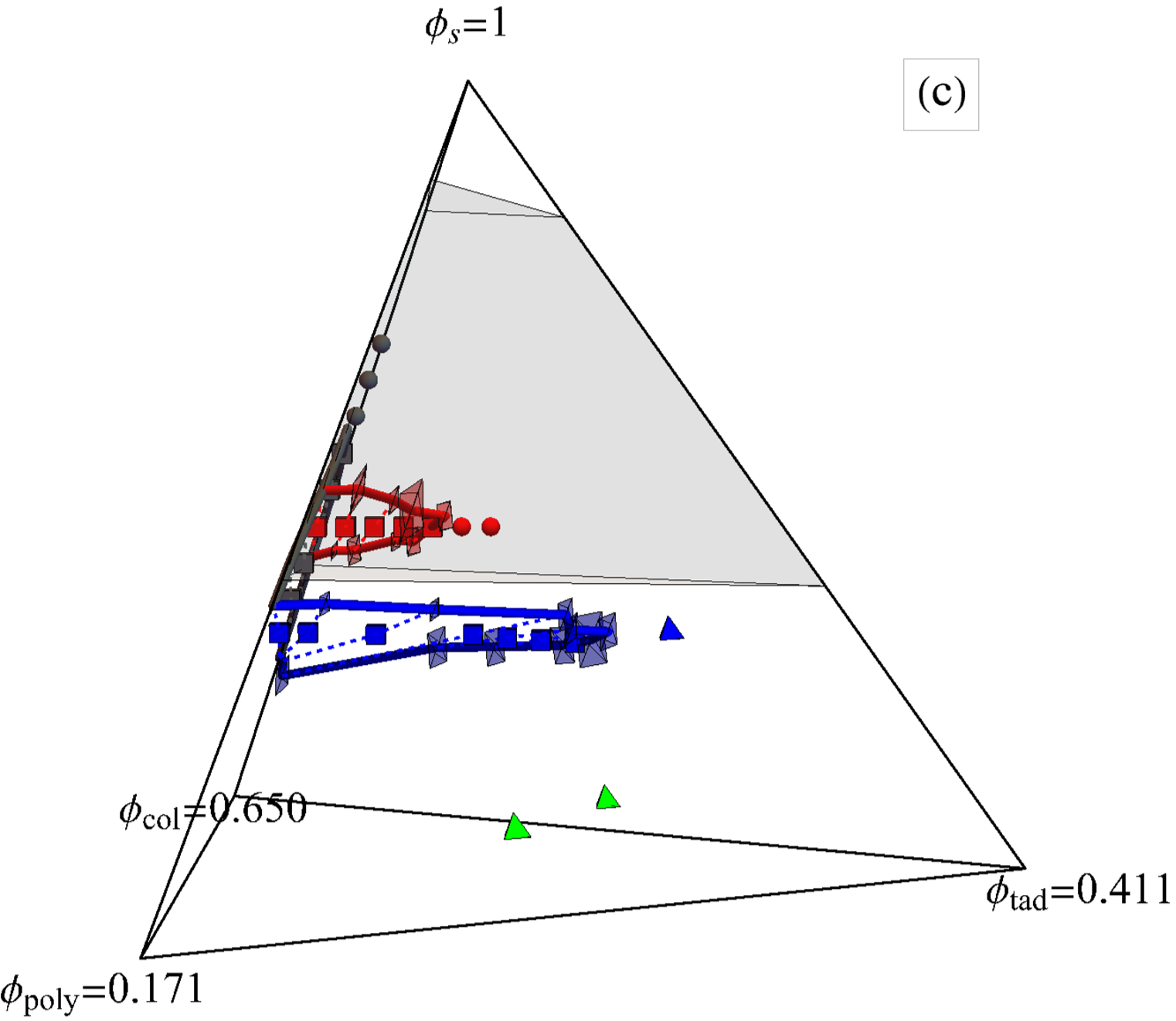}
    \includegraphics[width=0.49\linewidth]{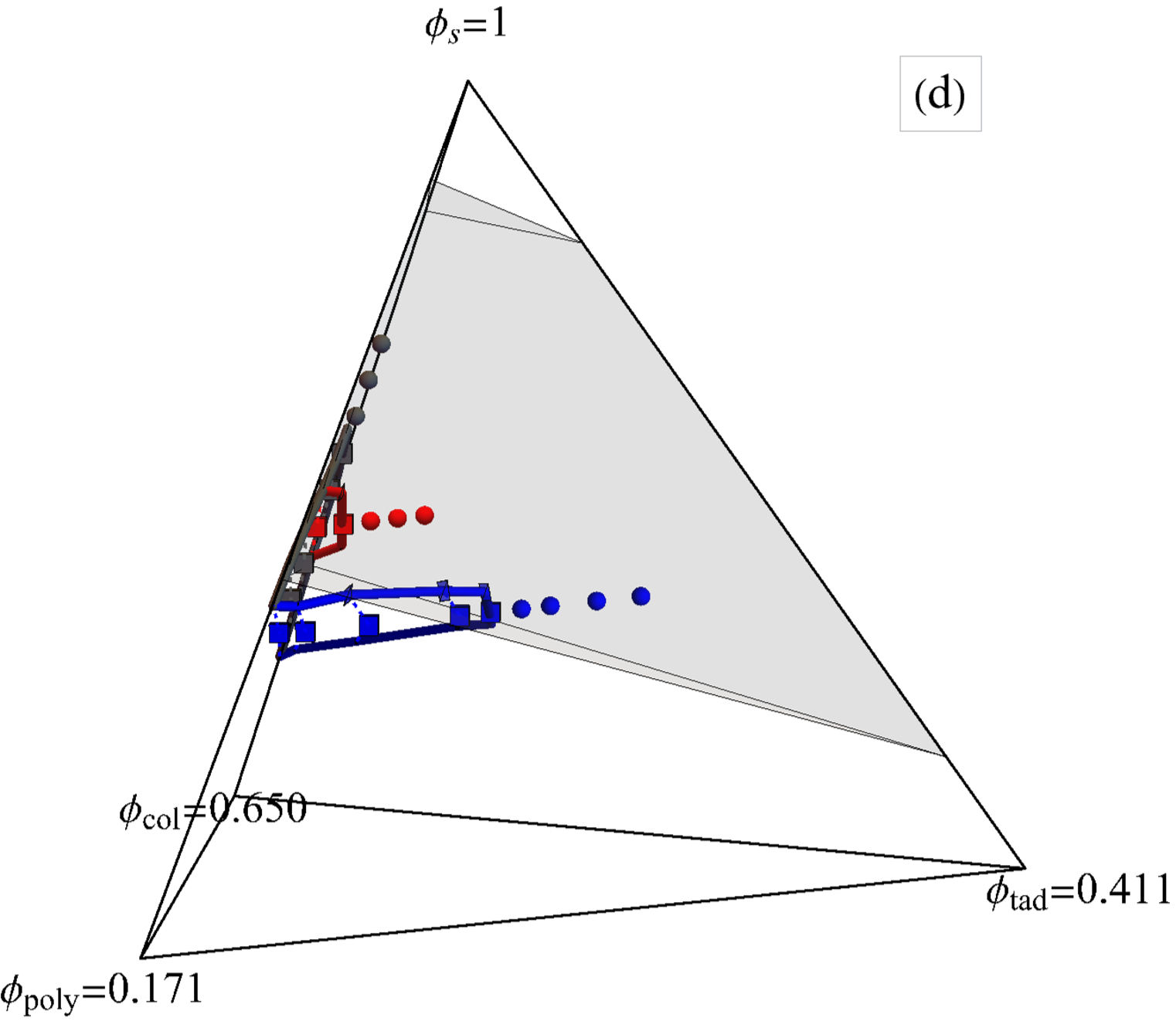}
    \caption{
    Phase diagrams for our model, with axes showing volume fractions for components - including the implicit solvent, $\phis$ - such that they sum to 1 everywhere inside each diagram.
    To better see the binodal region, the low-$\phis$ part of the phase diagrams has been omitted, and polymer and tadpole axes are shown on different scales, leading to the non-equal volume fractions at the corners.
    The end points of the dashed tie lines within data series have also been connected with lines of that series' colour.
    The key by marker shape indicates the stability of different points, and errors at the ends of tie lines are shown as diamonds whose lengths in each direction are twice the standard error on the mean.
    The key by colour identifies different  series of data, and both keys apply to all parts of the figure.
    The grey volume indicates the range of realistic osmotic pressures in Ref.~\cite{Mitchison2019}, mapped into model concentrations as discussed in Appendix \ref{Sec: SAFT}.
    Panel (a) shows the ternary phase diagram for systems with $\Ntad=0$, where contours are labeled to help read the $\phi$ values of our data.
    Panel (b) shows the 3D phase diagram including tadpoles with $\Ntail=30$, with 
    the 2D diagram in (a)  on its back-right face.
    Panel (c) shows a different view of (b)  showing how the binodals close as tadpoles are added.
    Panel (d) is the equivalent of (c) for systems with $\Ntail=15$.} 
    \label{Fig: phase diagram}
\end{figure*}

We have constructed a phase diagram for our model in Fig.~\ref{Fig: phase diagram}, using species volume fractions $\phi$ as axes.
For the volume of each species, we sum the volumes of individual particles within them, multiplied by the number of those molecules present.
Since the polymer pervaded volume is more important than its bare volume fraction for phase behaviour, its volume fraction is shown on different scale from the colloid to improve the clarity of results; similar applies for the tadpoles. 
Note this shears the contours in our ternary plot, which are shown in Fig.~\ref{Fig: phase diagram}(a).
Also, because our systems are compressible with varying total volume occupied by our particles, the implicit solvent needs including as its own component, hence including the solvent volume fraction $\phis = 1 - \phic - \phit - \phip$.

Our primary motivation is to determine where these systems undergo macrophase separation, as is believed to happen inside the cell nucleus and cytoplasm.
To that end, we have run several series of simulations as per the previous section, from which we map the binodal volume from the tie lines of macrophase-separating states.
The end points of these tie lines were found by measuring the volume fractions in each bulk phase, and averaging this over our independent simulations.
We can then compare this to the range of biologically relevant concentrations \cite{Mitchison2019}, shown as the shaded volume calculated using the Statistical Associating Fluid Theory (SAFT) formalism detailed in Appendix \ref{Sec: SAFT} \cite{Chapman1988,Chapman1990,Banaszak1996,Galindo1998}.
We use this formalism as a way to estimate the contribution of bonded chains in the pressure, which is difficult to measure in our hard systems, making it challenging to compare pressures directly.
However, SAFT differs from our model in that it fixes bond lengths to exactly $R_i + R_j$, and we found it over-predicts the number of our polymers required for a given pressure, as evidenced by the slightly different gradients of our tie lines and the shaded region's lower edge in Fig.~\ref{Fig: phase diagram}(a).
However, this will not alter our conclusions.

It is again helpful to focus first on the colloid + polymer systems with no tadpoles present.
This face of the phase diagram is shown in Fig.~\ref{Fig: phase diagram}(a), in which the blue and red data respectively encompass the two systems shown in Fig.~\ref{Fig: CP snapshots}. 
This demonstrates quantitatively that as the density decreases (towards $\phis=1$), the number of colloids found in the polymer-rich phase increases, although the density of polymers found in the colloid-rich phase remains small until the systems are a low enough density to mix.
Since such systems are nothing new \cite{Ilett1995}, our main interest is in its overlap of its binodal and the shaded biological concentrations.
While there is indeed some overlap, we caution that it is confined to the high concentration (low $\phis$) edge, and note that the author of Ref.~\cite{Mitchison2019} believed the lower osmotic pressures to be the most accurate. 
We are therefore mindful that the overlap we see may only exist due to inaccurate measurements extending the shaded region further than it should.

Whether that is the case or not, we infer from the limited overlap that entropy is unlikely to be strong enough by itself to drive biological phase separation.
Nevertheless, it is strong enough that we believe it should not be neglected as a contributing factor, especially as it is independent of chemistry and thus part of the interaction in all systems.

Fig.~\ref{Fig: phase diagram}(b) shows the phase diagram produced when we add tadpoles as per Sec.~\ref{Sec: qualitative behaviour}.
The higher density series of systems (in blue) includes those shown in Fig.~\ref{Fig: CTP snapshots} and extends out from the blue data in Fig.~\ref{Fig: phase diagram}(a).
A second, lower density series likewise extends out from the red data.
This viewpoint is useful for visualising how these series `grow' the binodal out from the colloid + polymer diagram, ultimately closing when enough tadpoles are added in this way.
Further viewpoints are provided in the Supplementary Material to help picture this phase diagram in 3D space.

The best viewpoint for seeing how quickly the binodal closes upon addition of tadpoles is shown in Fig.~\ref{Fig: phase diagram}(c), where the tadpole volume fraction increases rightwards away from the colloid-polymer plane (now found at the left edge).
Here we see both series of simulations map out planes in the binodal that are viewed nearly edge-on and that run close to parallel to the lower boundary of the shaded volume.
Interpolating the binodal between our series shows us that adding tadpoles decreases the overlap with the shaded volume.
Hence, the addition of tadpoles does not help entropy drive phase separation, as compared to the polymer + colloid case.
It therefore requires some inter-molecular attraction to account for biological phase separation.

Finally, we note that the micelle-producing systems in Fig.~\ref{Fig: micelles} were significantly more dense than biological systems (see the green triangles near the base of Fig.~\ref{Fig: phase diagram}(c)).
So while this is important behaviour for characterising the general properties of our tadpoles, we do not expect it to have much biological relevance.

\subsection{Differences when $\Ntail = 15$}
\label{Sec: Ntail=15}

We now investigate what changes when reduce the length of our tadpole tails from 30 to 15 beads, which would biologically represent shorter IDRs, e.g.~the shorter CTDs of RNA Polymerase II in yeast as compared to the human variant \cite{Boehning2018}.
We keep the length of polymers unchanged, to represent an otherwise unchanged environment.
Consequently, the replacement of colloids and polymers for an equal number of tadpoles no longer keeps the total number of beads fixed, and nor does the osmotic pressure remain approximately constant.

The phase diagram for these shorter tails is shown in Fig.~\ref{Fig: phase diagram}(d).
In this we show data series starting from the same two colloid + polymer systems as before, and again more viewpoints are provided in the Supplementary Material.
We kept the same axes as in the $\Ntail =30$ phase diagram to aid in their comparison, but due to the shorter tails, more tadpoles are required to create the same pressure.
This is why the shaded volume now extends down to a larger value on the $\phit$ axis.
Our procedure for adding tadpoles now also \emph{increases} $\phis$, leading to the data series rising as tadpoles are added.
So, where the tie lines and the base of the shaded volume were nearly co-planar with $\Ntail =30$, they now pass through each other.
On its own, this would help bring the binodal into the shaded volume.

However, we also find that the binodal now closes at a lower tadpole density for both data series.
This shrinking of the binodal is the larger effect of the two, resulting in an overall reduction in the capacity for tadpoles with shorter tails to support phase separation, as would be expected from their smaller conformational entropy.
We acknowledge, however, that longer tails also provide more interactions sites when chemistry is accounted for, such that analogous differences in experiments cannot easily be attributed solely to differences in entropy.

Another major change with shorter tails is their shifting preferences for bulk phases.
Where the $\Ntail=30$ tadpoles had a strong preference for the polymer-rich phase, Fig.~\ref{Fig: phi vs x}(b) shows the concentration of tadpoles with $\Ntail=15$ is approximately equal in both bulk phases.
Tail length can therefore be utilised to tune a macromolecule's preference for the separating phases.
It also has the advantage of being biologically accessible through mutations that repeat polypeptide sequences, as is thought to have happened with the CTD of RNA Polymerase II, without changing chemical interactions in the process.

\subsection{Interfacial tensions}
\label{Sec: Interface tension}

We now seek to quantitatively confirm that the interfacial tension $\gamma$ in our systems decreases with the addition of tadpoles.
$\gamma$ is difficult to access with hard potentials due to having infinitely poor statistics of the infinite forces present, so for this section we soften our system by using the Weeks-Chandler-Anderson (WCA) potential for particle repulsion,
\begin{equation}
    U_{\textrm{W}}(s) = \left\lbrace 
        \begin{array}{ll}
            4 u_\textrm{W}
            \left( s^{-12} - s^{-6} + 
            \frac{1}{4} \right) & \mathrm{for}~ s<2^{1/6}
            \\
            0 & \mathrm{otherwise}
        \end{array}
    \right. 
\end{equation}
where $s=r/R_\textrm{W}$ scales the particle separation, $r$, by the length scale of the WCA potential, $R_\textrm{W}$.
Bond potentials are also softened by using finitely extensible non-linear elastic (FENE) bonds with energy
\begin{equation}
    U_{\textrm{F}}(r) = \left\lbrace 
        \begin{array}{ll}
            -\frac{1}{2}k_{\textrm{F}} R_{\textrm{F}}^2
            \ln\! \left[ 1- \left(\dfrac{r}{R_{\textrm{F}}}\right)^2\right]  & \mathrm{for}~ r<R_{\textrm{F}}
            \\
            \infty & \mathrm{otherwise}
        \end{array}
    \right. .
\end{equation}
The new parameters were set to $u_\textrm{W} = \kT=\varepsilon$, $R_{\textrm{W}} = R_i + R_j$, $R_{\textrm{F}} = R_i + R_j + 4\Rtail$ between particles of species $i$ and $j$.
To match mean bond lengths to the hard systems, we set $k_\textrm{F} = 7.822\varepsilon/\ell^2$ between pairs of tail/polymer beads, and $k_\textrm{F} = 2.558\varepsilon/\ell^2$ between tadpole heads and tails.

With these potentials we have well-defined forces that can be calculated from system configurations.
This in turn lets us calculate the stress tensor $\bm\sigma$ and hence $\gamma$ via the Kirkwood-Buff expression \cite{Kirkwood1949}
\begin{equation}
    \gamma = -\frac{1}{2}\int\limits_{0}^{L_z}
        dz \left( \sigma_{zz} - \frac{1}{2}(\sigma_{xx}+\sigma_{yy})\right),
\label{Eq: Kirkwood-Buff}
\end{equation}
where having 2 interfaces in the system contributes the front factor of 1/2.

Finally, we change the box dimensions to $6 \times 6 \times 12$, which preserves the volume and hence particle numbers, but increases the ratio of interface area to bulk volume, improving our statistics.
In these shorted boxes we ran 3 sets of simulations equivalent to those in the higher density series in Sec.~\ref{Sec: qualitative behaviour} fusing colloids and polymers into tadpoles with  $\Ntail=30$.
To improve our statistics we ran these for longer to generate 200 independent snapshots, and in which we measured the interfacial tensions in Table \ref{Table: inteface tensions}.
These show a clear reduction in $\gamma$ with the addition of tadpoles, as expected if they act as surfactants.
These absolute values are in line with the commonly used estimate $\gamma \approx \kT/\delta^2$, with $\delta$ being the interface width.
We can also compare to the experimental values of 1 to $5\mu\mathrm{Nm}^{-1}$ measured by Jawerth et al.~\cite{Jawerth2018}, which overlaps with our range when we convert our units to SI, as shown in the fourth column of Table \ref{Table: inteface tensions}.

\begin{table}
\caption{\label{Table: inteface tensions} 
Interfacial tensions measured using Eq.~\ref{Eq: Kirkwood-Buff} in our softened systems.
Values in the third column have been scaled by the estimated interfacial width
$\delta = 2\Rcol(1 + q)$.
In the fourth column we have converted to SI units with $T$=300K, and $\delta = 36.3$nm.} 
\begin{tabular}{cccc}
$\Ntad$ & $\Ntad / \Npoly$ & $\gamma \; \delta^2 / \kT$ & $\gamma \:\: (\mu \mathrm{Nm}^{-1})$ \\
\hline
0 & 0 & $ 9.2 \pm 1.0 $ & $28.9\pm3$\\ 
47 & 1/3 & $ 4.1 \pm 1.1 $ & $12.9\pm3.5$ \\ 
94 & 1 & $ 0.4 \pm 1.0 $ & $1.3\pm3.1$ 
\end{tabular}
\end{table}

\section{Conclusions}

We have used minimal, equilibrium Monte Carlo simulations of hard sphere systems to investigate the role of entropy in biological liquid-liquid phase separation.
By constructing phase diagrams for solutions of polymers, colloids and tadpole molecules representing biomolecules with different entropic properties, we were able to compare the concentrations at which such systems are unstable to macrophase separation with concentrations found in real cell nuclei.
In doing so, we found entropy to be able to drive phase separation only at the upper estimates of biological concentrations.
We conclude that the entropy from macromolecules is a significant contribution, but insufficient without some further 
attractions.

An important agenda for further work would be to assess the source, strength and impact of attractive macromolecular interactions, and whether that leaves them weak enough for IDRs to remain flexible. 
Within the implicit solvent paradigm, this requires intermolecular interactions net of solvent effects, and the balance of exposed hydrophobic vs hydrophilic residues is a possible starting point. 
Beyond this there are explicit solvent effects to be considered, including solvent entropy.  
The latter has been argued to be a significant driver of some LLPS, though its relevance where the solvent is not excluded from one phase is unclear \cite{Pliss2010,Pliss2015,Galganski2017}.

Our simulations with tadpoles identified these molecules as entropic surfactants, exhibiting several typical surfactant behaviours: microphase separations such as micelles and bicontinuous phases; a preference for interfaces; and a reduction in interfacial tensions as tadpoles are added.
If such behaviour is present \textit{in vivo}, it would be a major step forwards in understanding the biological importance of IDRs and their influence on liquid drop formation and structure.
Indeed, if they exhibit strong segregation to the liquid drop interface, then the structure of `membraneless organelles' might be closer to that of regular organelles than was previously realised.

Another avenue for further work could be to model IDR-containing proteins with different morphologies, such as a ``pearl necklace" model for IDRs connected to globular regions at both ends.
We speculate that pearl necklace molecules would make less effective surfactants since both ends have low conformational entropy and would prefer the colloid-rich phase. 
However, this will be sensitive to the length of the IDR, which could tune the preference for the two phases and the interface like the length our tadpoles' tails does. 

Finally, in this article we simplified our model by matching particle sizes.
There remains the task of exploring parameter space where particle sizes are not matched, which will be an important step towards obtaining quantitative results for comparison with experiment.

\section*{Authors' contributions}
O.T.D.: data curation, formal analysis, investigation, methodology, software, visualisation, writing - original draft, and writing - review \& editing.
R.C.B.: conceptualisation, formal analysis, funding acquisition, supervision, and writing - review \& editing.

\section*{Funding}
This work was funded by EPSRC (Grant number: EP/T002794/1).

\section*{Acknowledgements}
We thank Dan Hebenstreit and Juntai Liu, and all the research team associated with that grant, for many discussions throughout the project.
We acknowledge the Warwick Scientific Computing RTP for providing the computational resources used.

\section*{Data accessibility}
The simulation and analysis codes can be found along with all data in the repository: http://wrap.warwick.ac.uk/163395/

\appendix 
\section{Verifying equilibration}
\label{App: verifying equilibration}

To determine when systems have equilibrated, we first calculate the radial distribution functions between all species $i$ and $j$, at each time $t$: $g_{ij}(r,t)$.
When doing this, we exclude pairs of particles within the same molecule.

\begin{figure*}
    \centering
    \includegraphics[width=0.49\linewidth]{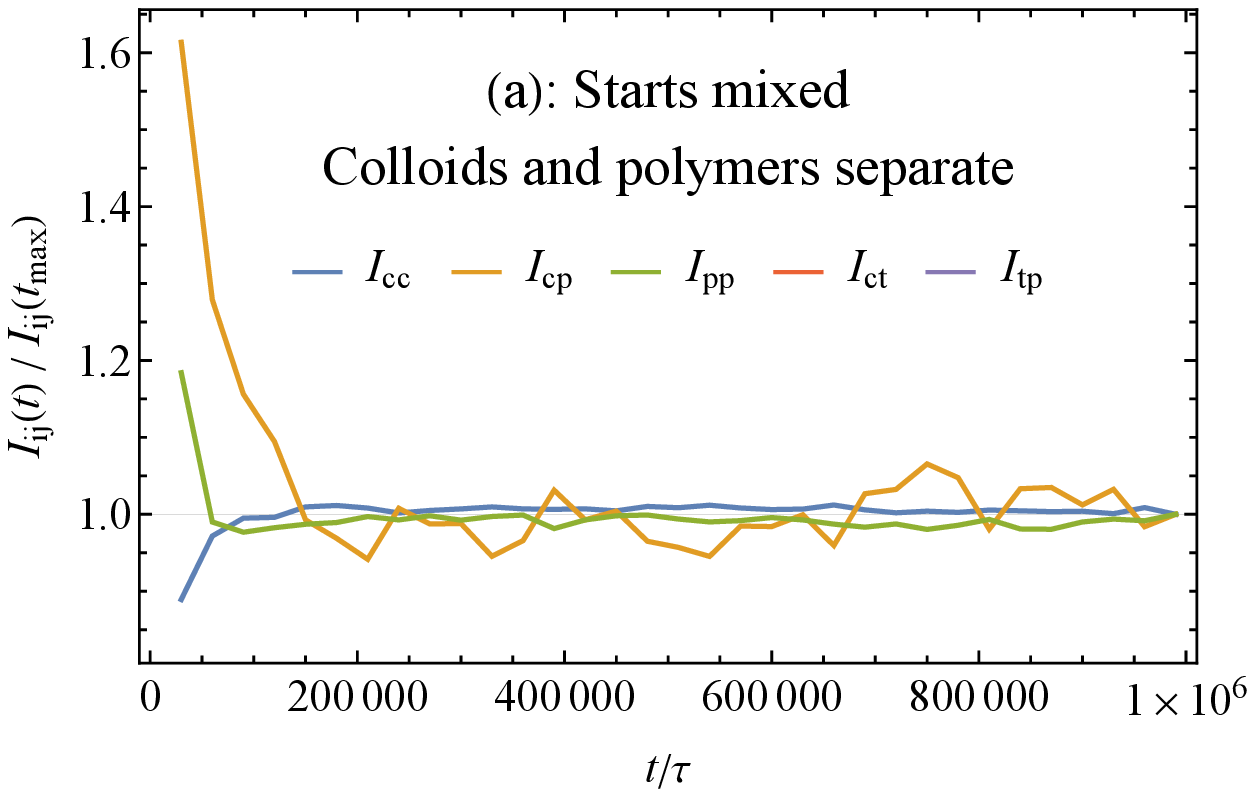}
    \includegraphics[width=0.49\linewidth]{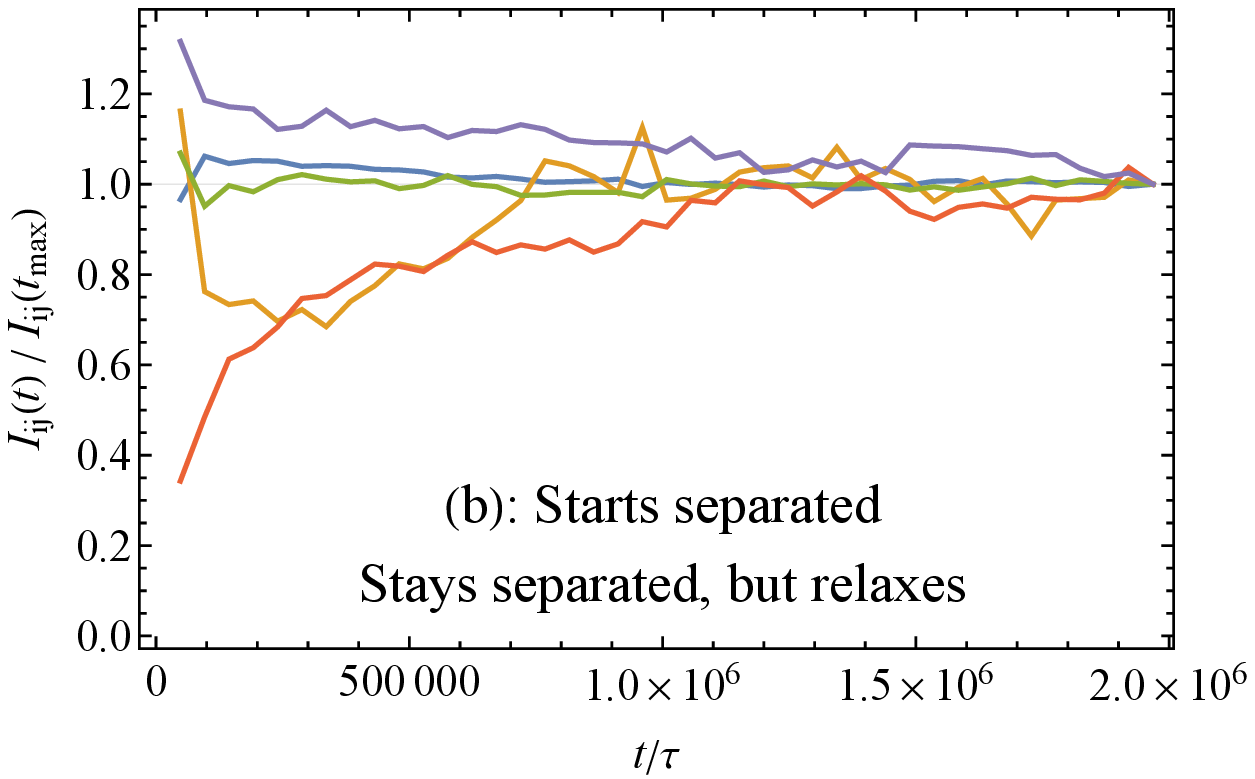}
    \includegraphics[width=0.49\linewidth]{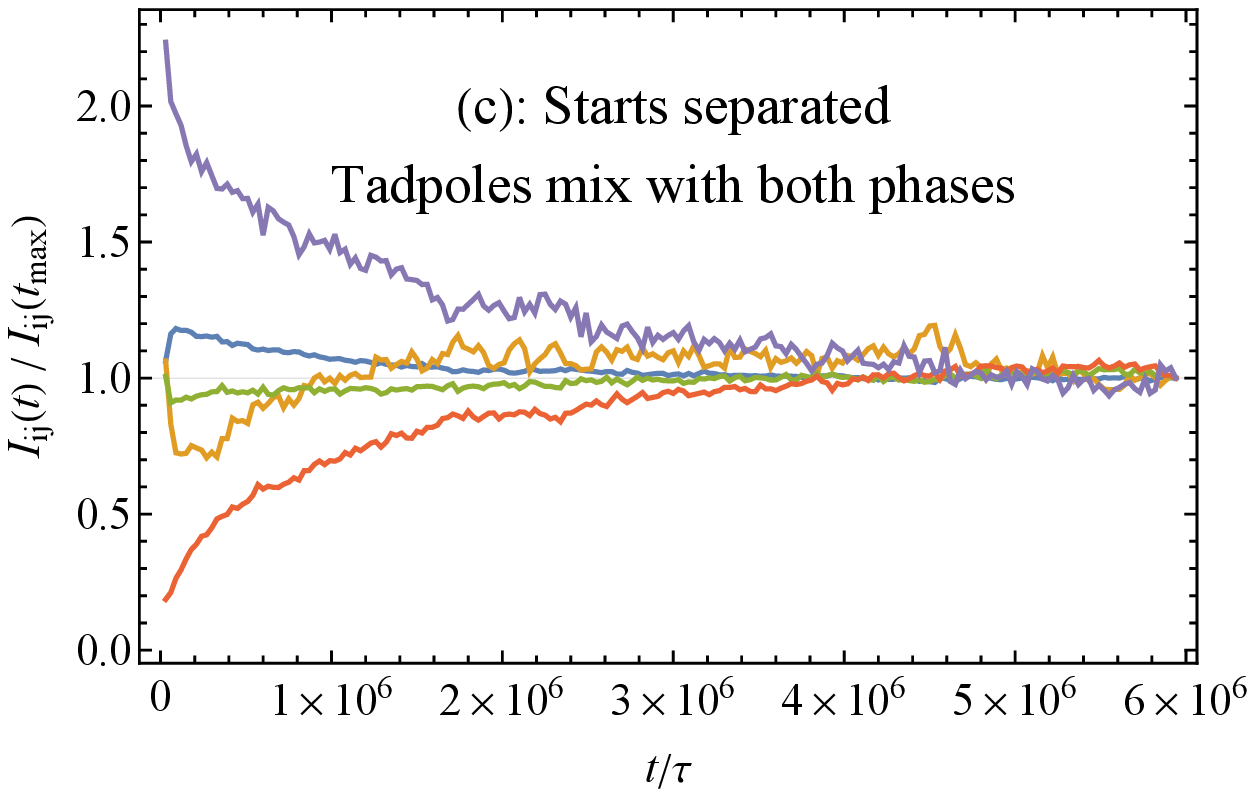}
    \includegraphics[width=0.49\linewidth]{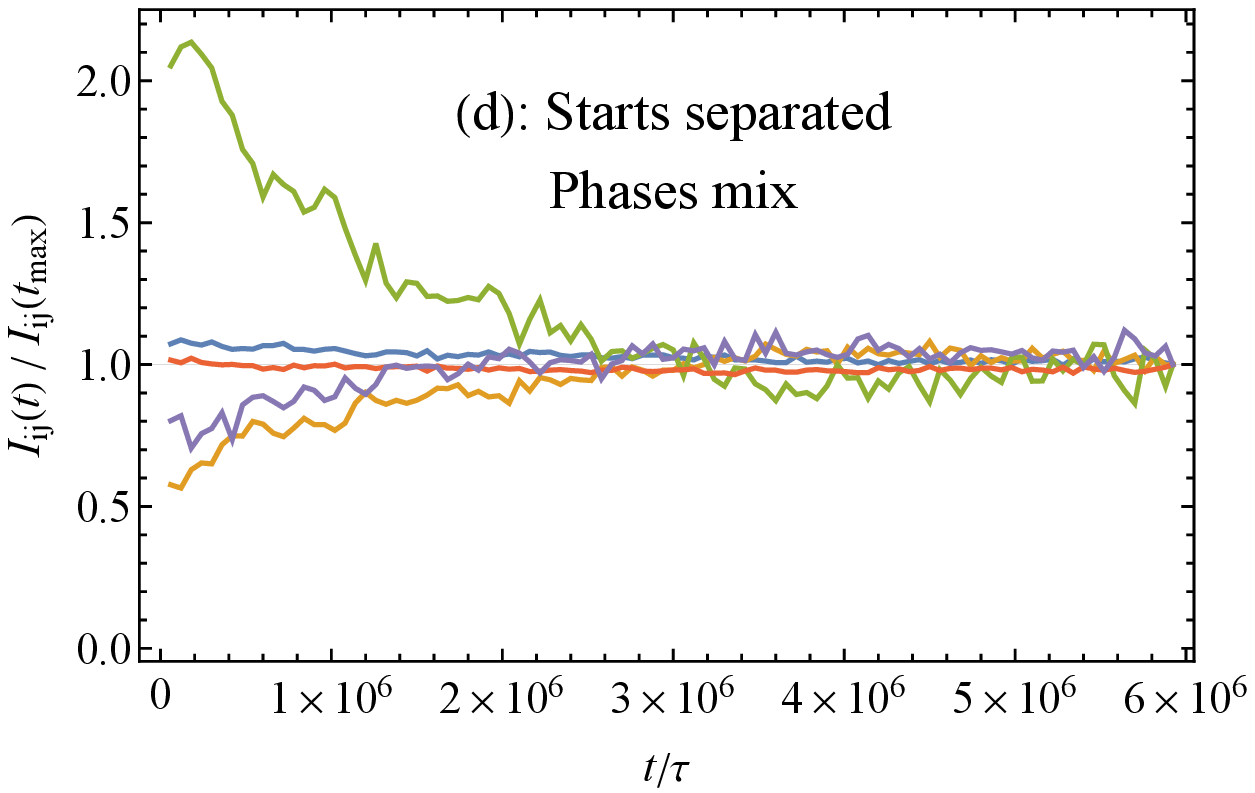}
    \caption{Plots of $I_{ij}(t)$, normalised by their value at the final time, showing the relaxation of cross-species radial distribution functions.
    The legend in part (a) applies to all plots, and the subscripts correspond to c=colloid, p=polymer and t=tadpole species.
    The initial system in (a) was mixed, while the other systems were initialised as separated states.
    The numbers of molecules in each panel were: (a): $\Ncol=374,~\Npoly=188,~\Ntad=0$;
    (b) and (c): $\Ncol=329,~\Npoly=141,~\Ntad=47$;
    (d): $\Ncol=235,~\Npoly=47,~\Ntad=141$.
    $\Ntail=30$ in (a) and (b), and $\Ntail=15$ in (c) and (d).}
    \label{Fig: Equilibration}
\end{figure*}

We then integrate these like so
\begin{equation}
    I_{ij}(t) = \int\limits_{0}^{2(R_{i}+R_{j})} 
        dr\, r^2 g_{ij}(r,t),
\end{equation}
where the upper limit roughly corresponds to the radius of the first minimum in a typical liquid RDF.
$I_{ij}(t)$ then quantifies the relaxation of the RDF.
Fig.~\ref{Fig: Equilibration} shows examples of this for a variety of systems evolving between different mixed and separated states.
In all plots we show $I_{\mathrm{cc}},~I_{\mathrm{pp}}$ and $I_{\mathrm{cp}}$ to show the relaxation of the two bulk phases.
Where tadpoles are present, we also show $I_{\mathrm{ct}}$ and $I_{\mathrm{tp}}$ to characterise their relaxation with respect to both phases.

We used Fig.~\ref{Fig: Equilibration}(a) to confirm our systems relax to separated states, and generated pre-separated states for systems containing tadpoles.
In Figs.~\ref{Fig: Equilibration}(b) and (c), which correspond to systems shown in Fig.~\ref{Fig: phi vs x}, the tadpoles started in the polymer phase where they mostly stayed in (b) though they do prefer the interface leading to a notable increase in $I_{ct}$ over time.
In part (c), many tadpoles leave the polymer phase, leading to a larger increase in $I_{\mathrm{ct}}$ and decrease in $I_{\mathrm{tp}}$.
The high densities made this a slow process, but exponential fits to our relaxation curves show we simulated out to about 6 times the relaxation time, making changes small enough at the end of the time series to treat as equilibrated.

In Fig.~\ref{Fig: Equilibration}(d) the tadpoles started with uniform density and the minority polymer component mixes with the colloid phase over time.
This is seen by $I_{\mathrm{pp}}$ relaxing from a large value as polymers lose sight of each other.

\section{The SAFT formalism}
\label{Sec: SAFT}

Statistical associating fluid theory (SAFT) provides a formalism for estimating the free energy in systems of associating particles, including polymers.
Here we summarise the SAFT expressions in Ref.~\cite{Galindo1998}, using our hard sphere potentials, starting by breaking the Helmholtz free energy is into 4 terms 
\begin{equation}
    F_{\mathrm{SAFT}} = F_{\mathrm{ideal}} + F_{\mathrm{part}} + F_{\mathrm{chain}} + F_{\mathrm{assoc}}.
\end{equation}
These respectively represent the ideal gas term, the contribution from individual particles in the absence of bonds, the contribution from intra-chain interactions, and the contribution from inter-chain interactions.

For our hard systems $F_{\mathrm{assoc}}=0$, while the ideal gas term is the usual sum over molecule species
\begin{equation}
    F_{\mathrm{ideal}} = \kT \sum\limits_{i} N_i \left(\ln\left( \frac{N_i}{V \lambda_{\mathrm{th},i}^3} \right) - 1\right),
\end{equation}
with the system volume $V$ and the thermal wavelength $\lambda_{\mathrm{th}}$.

The particle term uses an average contribution per particle calculated as
\begin{align}
\nonumber
    f_{\mathrm{part}} = \frac{1}{\zeta_0} &\left(
    \left( \frac{\zeta_{2}^3}{\zeta_{3}^2}     - \zeta_{0}\right)\ln(1-\zeta_{3}) \right. \\ 
    & \left. +            \frac{3\zeta_{1}\zeta_{2}}{1-\zeta_{3}} +
    \frac{\zeta_{2}^3}{\zeta_{3}(1-\zeta_{3})^2} \right) ,
\end{align}
where 
\begin{equation}
    \zeta_{\alpha} = \frac{\pi}{6V}
        \sum\limits_{i} N_{i} \sigma_{i}^{\alpha}. 
\end{equation}
and $\sigma_i$ is the diameter of particles of species $i$.
Applying this to all particles gives
\begin{equation}
    F_{\mathrm{part}} = \kT (\Ncol + \Ntad(1+\Ntail) + \Npoly\Nmono) f_{\mathrm{part}}.
\end{equation}

Finally, the chain term accounts for covalent bonds holding connected bead centres a distance $(\sigma_i + \sigma_j)/2$ apart.
I.e., the hard spheres are in contact.
This is accounted for through the value of the radial distribution function at contact
\begin{equation}
    g_{ij} = \frac{1}{1-\zeta_{3}} +
        \frac{3B_{ij}}{(1-\zeta_{3})^2} +
        \frac{2B_{ij}^2}{(1-\zeta_{3})^3},
\end{equation}
in which 
\begin{equation}
    B_{ij} = \frac{\sigma_{i} \sigma_{j}}{\sigma_{i} + \sigma_{j}} \zeta_{2}.
\end{equation}
Then 
\begin{align}
    \nonumber
    F_{\mathrm{chain}} = & -\kT \Ntad (\ln g_{\mathrm{ht}} + (\Ntail-1)\ln g_{\mathrm{tt}} ) \\
    & \;\;\;- \kT \Npoly(\Nmono-1) \ln g_{\mathrm{pp}},
\end{align}
counting all the bonds between tadpole heads, h, tadpole tail beads, t, and polymer beads, p.

From the free energy expression, we get the (osmotic) pressure from $p_{\mathrm{SAFT}} = -\left(\partial F_{\mathrm{SAFT}} / \partial V\right)_{T,N}$.
In this work, we solved this numerically to identify combinations of molecule densities with pressure equal to the upper and lower bounds found experimentally \cite{Mitchison2019}.
This range is then shown as the shaded volume in Fig.~\ref{Fig: phase diagram}.

\bibliographystyle{unsrt}
\bibliography{LLPT_refs}

\begin{thebibliography}{10}

\bibitem{Hyman2014}
A.~A. Hyman, C.~A. Weber, and F.~J\"{u}licher.
\newblock Liquid-liquid phase separation in biology.
\newblock {\em Annual Review of Cell and Developmental Biology}, 30:39--58,
  2014.

\bibitem{Alberti2017}
S.~Alberti.
\newblock The wisdom of crowds: regulating cell function through condensed
  states of living matter.
\newblock {\em The Journal of Cell Science}, 130:2789--2796, 2017.

\bibitem{Shin2017}
Y.~Shin and C.~P. Brangwynne.
\newblock Liquid phase condensation in cell physiology and disease.
\newblock {\em Science}, 357(6357), 2017.

\bibitem{Sawyer2019}
I.~A. Sawyer, D.~Sturgill, and M.~Dundr.
\newblock Membraneless nuclear organelles and the search for phases within
  phases.
\newblock {\em WIREs RNA}, 10(2):e1514, 2019.

\bibitem{Olson2015}
M.~OJ Olson and M.~Dundr.
\newblock {\em Nucleolus: Structure and Function}, pages 1--9.
\newblock American Cancer Society, 2015.

\bibitem{Galganski2017}
L.~Galganski, M.~O. Urbanek, and W.~J. Krzyzosiak.
\newblock Liquid-liquid phase separation in biology.
\newblock {\em Nucleic Acids Research}, 45:10350--10368, 2017.

\bibitem{Boehning2018}
M.~Boehning, C.~Dugast-Darzacq, and M.~Rankovic et~al.
\newblock Rna polymerase ii clustering through carboxy-terminal domain phase
  separation.
\newblock {\em Nat. Struct. Mol. Biol.}, 25:833--840, 2018.

\bibitem{Boija2018}
A.~Boija, I.~A. Klein, B.~R. Sabari, T.~I. Lee, D.~J. Taatjes, and R.~A. Young.
\newblock Transcription factors activate genes through the phase-separation
  capacity of their activation domains.
\newblock {\em Cell}, 175(7):1842--1855, 2018.

\bibitem{Brangwynne2015}
C.~Brangwynne, P.~Tompa, and R.~Pappu.
\newblock Polymer physics of intracellular phase transitions.
\newblock {\em Nature Physics}, 11:899--904, 2015.

\bibitem{Fuxreiter2018}
M.~Fuxreiter.
\newblock Fuzziness in protein interactions-a historical perspective.
\newblock {\em Journal of Molecular Biology}, 430(16):2278--2287, 2018.
\newblock Intrinsically Disordered Proteins.

\bibitem{Marenduzzo2006}
D.~Marenduzzo, K.~Finan, and P.~R. Cook.
\newblock The depletion attraction: an underappreciated force driving cellular
  organization.
\newblock {\em Journal of Cell Biology}, 175(5):681--686, 12 2006.

\bibitem{Brackley2013}
C.~A. Brackley, S.~Taylor, A.~Papantonis, P.~R. Cook, and D.~Marenduzzo.
\newblock Nonspecific bridging-induced attraction drives clustering of
  dna-binding proteins and genome organization.
\newblock {\em Proceedings of the National Academy of Sciences},
  110(38):E3605--E3611, 2013.

\bibitem{CanalsHamann2013}
A.~Z. Canals-Hamann, R.~P. das Neves, J.~E. Reittie, C.~I\~{n}iguez, S.~Soneji,
  T.~Enver, V.~J. Buckle, and F.~J. Iborra.
\newblock A biophysical model for transcription factories.
\newblock {\em BMC Biophysics}, 6, 2013.

\bibitem{Cook2018}
P.~R. Cook and D.~Marenduzzo.
\newblock Transcription-driven genome organization: a model for chromosome
  structure and the regulation of gene expression tested through simulations.
\newblock {\em Nucleic Acids Research}, 46(19):9895--9906, 09 2018.

\bibitem{Oh2018}
I.~Oh, S.~Choi, Y.~Jung, and J.~S. Kim.
\newblock Entropic effect of macromolecular crowding enhances binding between
  nucleosome clutches in heterochromatin, but not in euchromatin.
\newblock {\em Scientific Reports}, 8:5469, 2018.

\bibitem{Matsuda2014}
H.~Matsuda, G.~G. Putzel, V.~Backman, and I.~Szleifer.
\newblock Macromolecular crowding as a regulator of gene transcription.
\newblock {\em Biophysical Journal}, 106(8):1801--1810, 04 2014.

\bibitem{Cho2012}
E.~J. Cho and J.~S. Kim.
\newblock {Crowding Effects on the Formation and Maintenance of Nuclear Bodies:
  Insights from Molecular-Dynamics Simulations of Simple Spherical Model
  Particles}.
\newblock {\em Biophysical Journal}, 103(3):424--433, 08 2012.

\bibitem{Bakshi2015}
S.~Bakshi, H.~Choi, and J.~C. Weisshaar.
\newblock The spatial biology of transcription and translation in rapidly
  growing escherichia coli.
\newblock {\em Frontiers in Microbiology}, 6:636, 2015.

\bibitem{Kaur2019}
T.~Kaur, I.~Alshareedah, W.~Wang, J.~Ngo, M.~M. Moosa, and P.~R. Banerjee.
\newblock Molecular crowding tunes material states of ribonucleoprotein
  condensates.
\newblock {\em Biomolecules}, 9(2), 2019.

\bibitem{Andre2020}
A.~A.~M. Andr\'{e} and E.~Spruijt.
\newblock Liquid-liquid phase separation in crowded environments.
\newblock {\em International Journal of Molecular Sciences}, 21(16), 2020.

\bibitem{Asakura1954}
S.~Asakura and F.~Oosawa.
\newblock On interaction between two bodies immersed in a solution of
  macromolecules.
\newblock {\em The Journal of Chemical Physics}, 22(7):1255--1256, 1954.

\bibitem{Lekkerkerker1992}
H.~N.~W. Lekkerkerker, W.~C.-K. Poon, P.~N. Pusey, A.~Stroobants, and P.~B
  Warren.
\newblock Phase behaviour of colloid + polymer mixtures.
\newblock {\em Europhysics Letters ({EPL})}, 20(6):559--564, nov 1992.

\bibitem{Ilett1995}
S.~M. Ilett, A.~Orrock, W.~C.~K. Poon, and P.~N. Pusey.
\newblock Phase behavior of a model colloid-polymer mixture.
\newblock {\em Phys. Rev. E}, 51:1344--1352, Feb 1995.

\bibitem{Bolhuis2003}
P.~G. Bolhuis, E.~J. Meijer, and A.~A. Louis.
\newblock Colloid-polymer mixtures in the protein limit.
\newblock {\em Phys. Rev. Lett.}, 90:068304, Feb 2003.

\bibitem{Davis2022}
L.~K. Davis, I.~J. Ford, and B.~W. Hoogenboom.
\newblock Crowding-induced phase separation of nuclear transport receptors in
  fg nucleoporin assemblies.
\newblock {\em eLife}, 11:e72627, jan 2022.

\bibitem{Tadros2013}
T.~Tadros.
\newblock {\em Steric Stabilization}, pages 1048--1049.
\newblock Springer Berlin Heidelberg, Berlin, Heidelberg, 2013.

\bibitem{Zhang2003}
Z.~Zhang, M.~A. Horsch, M.~H. Lamm, and S.~C. Glotzer.
\newblock Tethered nano building blocks: Toward a conceptual framework for
  nanoparticle self-assembly.
\newblock {\em Nano Letters}, 3(10):1341--1346, 2003.

\bibitem{Iacovella2005}
C.~R. Iacovella, M.~A. Horsch, Z.~Zhang, and S.~C. Glotzer.
\newblock Phase diagrams of self-assembled mono-tethered nanospheres from
  molecular simulation and comparison to surfactants.
\newblock {\em Langmuir}, 21(21):9488--9494, 2005.
\newblock PMID: 16207026.

\bibitem{Iacovella2008}
C.~R. Iacovella, M.~A. Horsch, and S.~C. Glotzer.
\newblock Local ordering of polymer-tethered nanospheres and nanorods and the
  stabilization of the double gyroid phase.
\newblock {\em The Journal of Chemical Physics}, 129(4):044902, 2008.

\bibitem{Iacovella2011}
C.~R. Iacovella, A.~S. Keys, and S.~C. Glotzer.
\newblock Self-assembly of soft-matter quasicrystals and their approximants.
\newblock {\em Proceedings of the National Academy of Sciences},
  108(52):20935--20940, 2011.

\bibitem{Meli2009}
L.~Meli, A.~Arceo, and P.~F. Green.
\newblock Control of the entropic interactions and phase behavior of athermal
  nanoparticle/homopolymer thin film mixtures.
\newblock {\em Soft Matter}, 5:533--537, 2009.

\bibitem{Kumar2013}
S.~K. Kumar, N.~Jouault, B.~Benicewicz, and T.~Neely.
\newblock Nanocomposites with polymer grafted nanoparticles.
\newblock {\em Macromolecules}, 46(9):3199--3214, 2013.

\bibitem{Marson2015}
R.~L. Marson, T.~D. Nguyen, and S.~C. Glotzer.
\newblock Rational design of nanomaterials from assembly and reconfigurability
  of polymer-tethered nanoparticles.
\newblock {\em MRS Communications}, 5(3):397--406, 2015.

\bibitem{Kumar2017}
S.~K. Kumar, V.~Ganesan, and R.~A. Riggleman.
\newblock Perspective: Outstanding theoretical questions in
  polymer-nanoparticle hybrids.
\newblock {\em The Journal of Chemical Physics}, 147(2):020901, 2017.

\bibitem{Matsarskaia2016}
O.~Matsarskaia, M.~K. Braun, F.~Roosen-Runge, M.~Wolf, F.~Zhang, R.~Roth, and
  F.~Schreiber.
\newblock Cation-induced hydration effects cause lower critical solution
  temperature behavior in protein solutions.
\newblock {\em The Journal of Physical Chemistry B}, 120(31):7731--7736, 2016.
\newblock PMID: 27414502.

\bibitem{Sahoo2022}
A.~K. Sahoo, F.~Schreiber, R.~R. Netz, and P.~K. Maiti.
\newblock Role of entropy in determining the phase behavior of protein
  solutions induced by multivalent ions.
\newblock {\em Soft Matter}, 18:592--601, 2022.

\bibitem{Bianco2019}
V.~Bianco, M.~Alonso-Navarro, D.~Di~Silvio, S.~Moya, A.~L. Cortajarena, and
  I.~Coluzza.
\newblock Proteins are solitary! pathways of protein folding and aggregation in
  protein mixtures.
\newblock {\em The Journal of Physical Chemistry Letters}, 10(17):4800--4804,
  2019.
\newblock PMID: 31373499.

\bibitem{Bianco2020}
V.~Bianco, G.~Franzese, and I.~Coluzza.
\newblock In silico evidence that protein unfolding is a precursor of protein
  aggregation.
\newblock {\em ChemPhysChem}, 21(5):377--384, 2020.

\bibitem{Park2020}
S.~Park, R.~Barnes, Y.~Lin, B.-J. Jeon, S.~Najafi, K.~T. Delaney, G.~H.
  Fredrickson, J.-E. Shea, D.~S. Hwang, and S.~Han.
\newblock Dehydration entropy drives liquid-liquid phase separation by
  molecular crowding.
\newblock {\em Communications Chemistry}, 3(1):83, 2020.

\bibitem{Allen2017}
M.~P. Allen and D.~J. Tildesley.
\newblock {\em {Computer Simulation of Liquids}}.
\newblock Oxford University Press, 2017.

\bibitem{Davis2002}
J.~A. Davis, Y.~Takagi, R.~D. Kornberg, and F.~J. Asturias.
\newblock Structure of the yeast rna polymerase ii holoenzyme: Mediator
  conformation and polymerase interaction.
\newblock {\em Molecular Cell}, 10:409--415, 2002.

\bibitem{Marques2013}
E.~F. Marques and B.~F.~B. Silva.
\newblock {\em Surfactants, Phase Behavior}, pages 1290--1333.
\newblock Springer Berlin Heidelberg, Berlin, Heidelberg, 2013.

\bibitem{Mitchison2019}
T.~J. Mitchison.
\newblock Colloid osmotic parameterization and measurement of subcellular
  crowding.
\newblock {\em Molecular biology of the cell}, 30(2):173--180, January 2019.

\bibitem{Chapman1988}
W.~G. Chapman, G.~Jackson, and K.~E. Gubbins.
\newblock Phase equilibria of associating fluids.
\newblock {\em Molecular Physics}, 65(5):1057--1079, 1988.

\bibitem{Chapman1990}
W.~G. Chapman, K.~E. Gubbins, G.~Jackson, and M.~Radosz.
\newblock New reference equation of state for associating liquids.
\newblock {\em Industrial and Engineering Chemistry Research},
  29(8):1709--1721, 1990.

\bibitem{Banaszak1996}
M.~Banaszak, C.~K. Chen, and M.~Radosz.
\newblock Copolymer saft equation of state. thermodynamic perturbation theory
  extended to heterobonded chains.
\newblock {\em Macromolecules}, 29(20):6481--6486, 1996.

\bibitem{Galindo1998}
A.~Galindo, L.~A. Davies, A.~Gil-Villegas, and G.~Jackson.
\newblock The thermodynamics of mixtures and the corresponding mixing rules in
  the saft-vr approach for potentials of variable range.
\newblock {\em Molecular Physics}, 93(2):241--252, 1998.

\bibitem{Kirkwood1949}
J.~G. Kirkwood and F.~P. Buff.
\newblock The statistical mechanical theory of surface tension.
\newblock {\em The Journal of Chemical Physics}, 17(3):338--343, 1949.

\bibitem{Jawerth2018}
L.~M. Jawerth, M.~Ijavi, M.~Ruer, S.~Saha, M.~Jahnel, A.~A. Hyman,
  F.~J\"ulicher, and E.~Fischer-Friedrich.
\newblock Salt-dependent rheology and surface tension of protein condensates
  using optical traps.
\newblock {\em Phys. Rev. Lett.}, 121:258101, Dec 2018.

\bibitem{Pliss2010}
A.~Pliss, A.~N. Kuzmin, A.~V. Kachynski, and P.~N. Prasad.
\newblock Nonlinear optical imaging and raman microspectrometry of the cell
  nucleus throughout the cell cycle.
\newblock {\em Biophysical Journal}, 99:3483--3491, 2010.

\bibitem{Pliss2015}
A.~Pliss, X.~Peng, L.~Liu, A.~Kuzmin, Y.~Wang, J.~Qu, Y.~Li, and P.~N. Prasad.
\newblock Single cell assay for molecular diagnostics and medicine: Monitoring
  intracellular concentrations of macromolecules by two-photon fluorescence
  lifetime imaging.
\newblock {\em Theranostics}, 5:919--930, 2015.

\end{thebibliography}

\end{document}